\documentclass[pdflatex,sn-mathphys-num,iicol]{sn-jnl}

\usepackage[mathlines]{lineno}
\usepackage{graphicx}%
\usepackage{caption}
\usepackage{multirow}%
\usepackage{amsmath,amssymb,amsfonts}%
\usepackage{amsthm}%
\usepackage{mathrsfs}%
\usepackage[title]{appendix}%
\usepackage{xcolor}%
\usepackage{textcomp}%
\usepackage{manyfoot}%
\usepackage{booktabs}%
\usepackage{algorithm}%
\usepackage{algorithmicx}%
\usepackage{algpseudocode}%
\usepackage{listings}%
\usepackage{siunitx}
\usepackage{chngcntr}

\unnumbered

\newcommand{\Vrel}{\mathcal{V}}
\newcommand{\Arel}{\mathcal{A}}
\newcommand{\drel}{\mathcal{D}}

\begin{document}

\title[Ultrafast X-ray sonography]{Ultrafast X-ray sonography reveals the spatial heterogeneity of the laser-induced magneto-structural phase transition in FeRh}

\author*[1]{\fnm{Maximilian} \sur{Mattern}}\email{mattern@mbi-berlin.de}
\author[2]{\fnm{Angel} \sur{Rodriguez-Fernandez}}
\author[2]{\fnm{Roman} \sur{Shayduk}}
\author[3]{\fnm{Jon Ander} \sur{Arregi}}
\author[3,4]{\fnm{Vojt\v{e}ch} \sur{Uhl\'{i}\v{r}}}
\author[2]{\fnm{Ulrike} \sur{Boesenberg}}
\author[2]{\fnm{Jörg} \sur{Hallmann}}
\author[2]{\fnm{Wonhyuk} \sur{Jo}}
\author[5,2]{\fnm{Aliaksandr} \sur{Leonau}}
\author[2]{\fnm{Rustam} \sur{Rysov}}
\author[2]{\fnm{James} \sur{Wrigley}}
\author[2]{\fnm{Alexey} \sur{Zozulya}}
\author[1,6]{\fnm{Stefan} \sur{Eisebitt}}
\author[2]{\fnm{Anders} \sur{Madsen}}
\author*[1]{\fnm{Daniel} \sur{Schick}}\email{schick@mbi-berlin.de}
\author*[2]{\fnm{Jan-Etienne} \sur{Pudell}}\email{jan-etienne.pudell@xfel.eu}

\affil[1]{\small\orgname{Max Born Institute for Nonlinear Optics and Short Pulse Spectroscopy}, \orgaddress{\city{Berlin}, \postcode{12489}, \country{Germany}}}

\affil[2]{\small\orgname{European X-ray Free-Electron Laser Facility}, \orgaddress{\city{Schenefeld}, \postcode{22869}, \country{Germany}}}

\affil[3]{\small\orgdiv{CEITEC BUT}, \orgname{Brno University of Technology}, \orgaddress{ \city{Brno}, \postcode{61200}, \country{Czech Republic}}}

\affil[4]{\small\orgdiv{Institute of Physical Engineering}, \orgname{Brno University of Technology}, \orgaddress{\city{Brno}, \postcode{61200}, \country{Czech Republic}}}

\affil[5]{\small\orgdiv{Department Physik}, \orgname{Universität Siegen}, \orgaddress{\city{Siegen}, \postcode{57072}, \country{Germany}}}

\affil[6]{\small\orgdiv{Institut für Physik und Astronomie}, \orgname{Technische Universität Berlin}, \orgaddress{\city{Berlin}, \postcode{10623}, \country{Germany}}}

\abstract{
Phase transitions are governed by both intrinsic and extrinsic heterogeneities, yet capturing their spatio-temporal dynamics remains a challenge.
While ultrafast techniques track phase changes on femtosecond timescales, the spatial complexity and stochastic nature of the processes often remain hidden.
Here, we present an experimental approach that combines well-established ultrafast hard-X-ray diffraction with a propagating strain pulse as a universal and non-invasive probe.
This ultrafast X-ray sonography can capture the spatio-temporal phase heterogeneity in great detail by resolving the phase-specific strain response.
We apply this approach to the antiferromagnetic-to-ferromagnetic magneto-structural phase transition in FeRh and identify the ferromagnetic phase to nucleate at the surface as narrow columnar domains of approximately \qty{30}{nm} diameter. 
Besides reconciling the diverse experimental results in the literature on FeRh, X-ray sonography offers a versatile platform for investigating a wide range of phase transitions accompanied by structural changes.
}


\maketitle

\section{Introduction}
Heterogeneity is a hallmark of first-order phase transitions, where long-range order emerges from the interplay and coexistence of competing phases~\cite{dagotto2005,seul1995,vidas2018,mcleod2017,zhang2002,pressacco2016}.
These spatial variations, ranging from static defects~\cite{ahn2022,monti2024,ocallahan2015} to transient fluctuations~\cite{zong2019,lee2012}, critically influence the microscopic nature and kinetics of the phase transition and hence also the resulting macroscopic properties of the material.
The advent of ultrafast scattering techniques has been instrumental in revealing the complex interplay of heterogeneous electronic (charge, spin, and orbital) and structural dynamics in a series of ground-breaking experiments on ultrafast phase transitions with femtosecond to picosecond temporal and nanometre spatial resolution~\cite{ahn2022,ocallahan2015,johnson2023,johnson2024,forst2015,danz2021}. 
Especially in technologically relevant, nanoscale heterostructures, symmetry breaking at interfaces and vastly different in-plane vs. out-of-plane dimensions cause an additional anisotropy of the spatio-temporal dynamics~\cite{ahn2022,amano2024}.
Also the intrinsically heterogeneous ultrafast drivers of the phase transition contribute to the anisotropy of the dynamics~\cite{mariager2012,forst2015,mattern2024b,monti2024}.
Capturing this \emph{directional heterogeneity} is, however, essential for unravelling key processes during ultrafast phase transitions.To address this challenge, we introduce \emph{ultrafast X-ray sonography} --- a combination of time-resolved hard-X-ray diffraction with sound waves in form of high-frequency acoustic strain pulses of picosecond duration.
We leverage the strain pulses as a transient structural marker, enabling localisation of coexisting phases in both time and space by following their evolving X-ray signatures on ultrafast time scales.
This non-invasive pump-probe technique can track the coherent growth and stochastic nucleation of structural domains with almost no requirements for sample preparation. 
Our approach is broadly applicable to a wide class of materials, provided that the coexisting phases exhibit distinguishable diffraction signatures~\cite{randi2016,cavalleri2001,mcleod2017,gatel2017,mariager2012,baum2007,jong2013,gedik2007}.

We demonstrate the power of ultrafast X-ray sonography by applying it to the prototypical antiferromagnetic-to-ferromagnetic (AFM–FM) magneto-structural phase transition in FeRh.
Our measurements reveal that the FM phase nucleates at the surface in narrow, vertically extended columnar domains with a diameter of \qty{30}{nm}, hence, directly visualising a process previously obscured in spatially averaging experiments~\cite{li2022,mattern2024a}. 
These findings provide a common ground for interpreting the diverse experimental observations on FeRh~\cite{li2022,mariager2012,mattern2024a} and underscore the importance of directional heterogeneities governing the kinetics of first-order phase transitions.

\section{Ultrafast X-ray sonography}

\begin{figure*}[t]
\centering
\includegraphics[width = \textwidth]{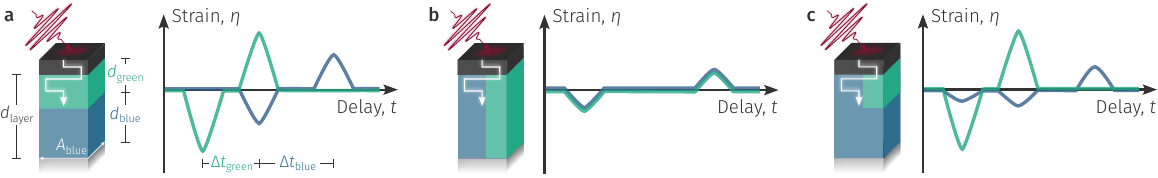}
\caption{\label{fig:fig_1_idea} \textbf{Concept of ultrafast X-ray sonography:}
The phase-specific strain response of two structural phases (blue and green) in case of \textbf{a}, out-of-plane, \textbf{b}, lateral, and \textbf{c}, lateral and out-of-plane separation within a layer of thickness $d_\text{layer}$.
The heterogeneity can be parametrised by the thickness $d_\text{green/blue}$ and covered area $A_\text{green/blue}$ of the two phases.
The transducer (dark grey) launches a bipolar strain pulse (white solid arrow) that propagates through the layer towards the substrate (bright grey), thereby introducing a transient strain of both phases:
\textbf{a},
The bipolar strain pulse first enters the upper and later the lower phase-separated region, leading to qualitatively similar, but delayed strain responses for both phases.
\textbf{b},
The bipolar strain pulse traverses both phases simultaneously, and their strain responses are identical.
\textbf{c},
The shape of the strain response differs between the two phases as the strain response of the phase that is only present in the region of lateral coexistence (green) is identical to case \textbf{a} and the strain response of the other phase (blue) is the superposition of case \textbf{a} and \textbf{b}.
}
\end{figure*}

Propagating sound waves transiently strain materials and modify their structure.
In case of a bipolar strain pulse, the average strain of a layer, however, vanishes when both the compressive and expansive part are located within the layer.
Only the entrance and exit of the strain pulse, when the compressive and expansive part within the layer are unbalanced, leads on average to a compression (negative strain) or expansion (positive strain) of the layer.
Accordingly, when the centre of the strain pulse reaches the top and bottom interface, the layer exhibits the maximum compression and expansion, respectively.
In ultrafast X-ray diffraction (UXRD) experiments, this transient average strain translates to a characteristic shift of the layer's structural Bragg peak~\cite{mattern2023a}.
In the case of the coexistence of two distinct structural phases, the specific strain response of each phase enables its direct localisation along the layer's depth and therefore provides access to spatial heterogeneities both in and out of the plane. 
This key concept of ultrafast X-ray sonography, namely the relation between phase heterogeneity and phase-specific strain response, is illustrated in Fig.~\ref{fig:fig_1_idea}.
For simplicity, we first focus on a static phase coexistence and assume identical sound velocities for both phases, neglecting reflections of the strain pulse from all interfaces.
With the help of a photoacoustic transducer (dark grey), a bipolar strain pulse is launched into the layer of thickness $d_\text{layer}$ hosting two structural phases (blue and green).
The resulting phase-specific strain response allows distinguishing three major scenarios for the phase separation, see panels~Fig.~\ref{fig:fig_1_idea}a-c: an exclusive out-of-plane phase separation (a), an exclusive in-plane phase separation (b) and a mixed in- and out-of-plane separation (c).

While regular UXRD can only provide the volume fractions, $\Vrel_\text{phase}(t)$, of the individual phases by the ratio of their corresponding integrated Bragg peak intensities~\cite{cavalleri2001,mcleod2017,mariager2012,baum2007,jong2013,gedik2007,mattern2022,amano2024}, ultrafast X-ray sonography further enables capturing the three-dimensional nature of the phase heterogeneity.
In addition to this qualitative insight, the delay between the prominent maxima and minima of the phase-specific strain response, $\Delta t_\text{phase}$, directly decodes the thickness of each phase, $d_\text{phase}$, via
\begin{equation}    
    d_\text{phase}=\frac{\Delta t_\text{phase}}{v_s} \;,
    \label{eq:eq_0_phase_thickness}
\end{equation}
with $v_s$ being the longitudinal sound velocity.
Similarly, the delay between the maximum compression of both phases quantifies the depth of the phase boundary within the layer.
To fully parametrise the spatio-temporal phase heterogeneity, we can determine the relative in-plane coverage of each phase, $\Arel_\text{phase}(t)=A_\text{phase}(t) / A_\text{layer}$, by relating it to $\Vrel_\text{phase}(t)$ and the relative phase thickness $\drel_\text{phase}(t)=d_\text{phase}(t) / d_\text{layer}$ via
\begin{equation}    
    \Vrel_\text{phase}(t)=\drel_\text{phase}(t)\ \Arel_\text{phase}(t) \; .
    \label{eq:eq_1_separation}
\end{equation}

\section{Sonogram of the laser-driven magneto-structural phase transition in {FeRh}}

To demonstrate and benchmark the capabilities of ultrafast X-ray sonography, we study the transient heterogeneity of the prototypical first-order AFM-FM magneto-structural phase transition in FeRh.
In thermal equilibrium, the transition at $\approx \qty{370}{K}$ is characterised by a change in the electronic band structure~\cite{gray2012, pressacco2018}, an arising FM magnetisation~\cite{stamm2008}, and an isotropic expansion of the unit cell~\cite{arregi2020}.
The latter enables tracking phase-specific strain responses, thereby facilitating the application of ultrafast X-ray sonography to probe the spatio-temporal appearance of local FM order.

Previous UXRD experiments focused on the timescale of the emerging FM phase as determined by the rise of its structural Bragg peak~\cite{mariager2012,mattern2024a,mattern2024b}. 
The observed single-exponential increase of the FM volume fraction, 
\begin{equation}
    \Vrel_\text{FM}(t) = \mathcal{H}(t)\, \Vrel_\text{FM}^* \cdot \left( 1-e^{-t/\tau} \right) \ ,\label{eq:eq_0_model_old}
\end{equation}
is consistent with a stochastic nucleation of FM domains at independent sites, as described by an Arrhenius activation over an energy barrier~\cite{avrami1939,mattern2025,dolgikh2025}. 
Here, $t$ denotes the pump-probe delay, $\mathcal{H}(t)$ the Heaviside step function, and $\Vrel_\text{FM}^*$ the final FM volume fraction.
The rapid modification of the electronic band structure of FeRh upon direct electron-photon interaction~\cite{pressacco2021} perturbs the critical Rh-Fe hybridisation~\cite{mattern2025,hama2024,kang2023}, resulting in a drastically reduced nucleation timescale of $\tau = \qty{8}{ps}$, as compared to \qty{50}{ps} upon near-equilibrium heating~\cite{mattern2024a,mattern2024b,mattern2025}.
We want to emphasise that only tracking $\Vrel_\text{FM}(t)$, as done so far, fails to provide any insights into the transient phase heterogeneity.

\begin{figure*}[t!]
\centering
\includegraphics[width = \textwidth]{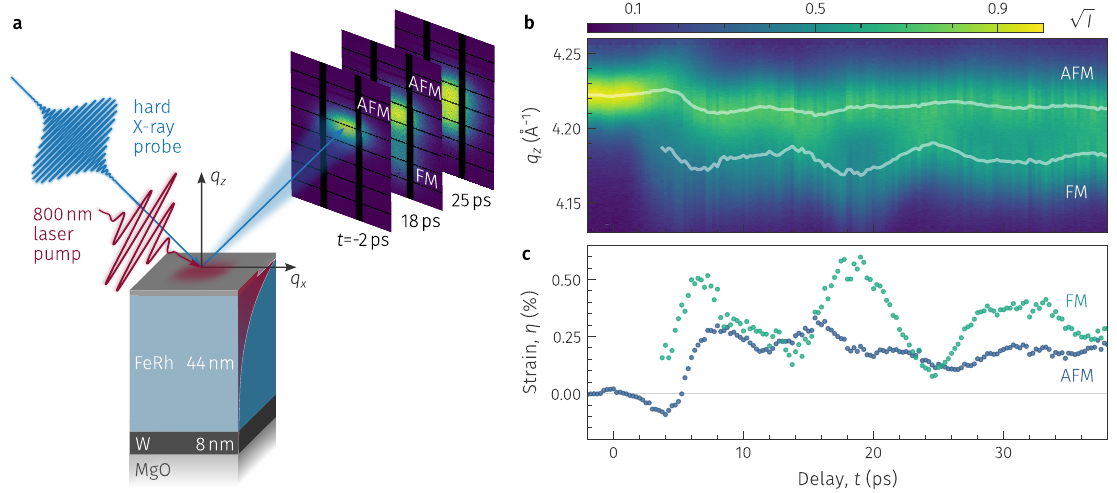}
\caption{\label{fig:fig_2_experiment} \textbf{Ultrafast X-ray sonography on FeRh:} 
\textbf{a}, Sketch of the pump-probe ultrafast X-ray diffraction experiment applying reciprocal space slicing~\cite{zeuschner2021} including cutouts of detector images at different pump-probe delays.
The optical absorption profile (red shade) is shown on the side face of the sample.
\textbf{b}, Temporal evolution of the normalised diffracted intensity along the out-of-plane reciprocal coordinate $q_z$ upon excitation with an incident optical laser fluence of \qty{7.7}{\mJ\per\cm\squared}. 
After excitation at $t = \qty{0}{ps}$, the structural Bragg peak of the ferromagnetic (FM) phase emerges at smaller $q_z$. 
The solid lines denote the transient positions of the antiferromagnetic (AFM) and FM Bragg peaks.
\textbf{c}, Extracted transient strain response of the AFM (blue) and FM (green) phases.
}
\end{figure*}

Utilising ultrafast X-ray sonography, we can now provide this important puzzle piece and determine $\Arel_\text{FM}(t)$ and $\drel_\text{FM}(t)$ through the non-equilibrium pathway of the phase transition in FeRh.
In contrast to the conceptual Fig.~\ref{fig:fig_1_idea}, the sample structure contains no dedicated transducer as source of the strain pulses and hence requires no adaptation for the experiment.
Instead, the direct optical excitation of FeRh (see Fig.~\ref{fig:fig_2_experiment}a) that unlocks the non-equilibrium pathway of the phase transition simultaneously launches the strain pulse probing the transient phase heterogeneity.
We follow the phase-specific lattice response after photo-excitation by femtosecond near-infrared laser pulses employing UXRD with monochromatic, femtosecond hard-X-ray pulses and a position-sensitive area detector provided at the Materials Imaging and Dynamics (MID) instrument~\cite{madsen2021} at the European X-ray Free-Electron Laser Facility (EuXFEL). 
More details on the sample, experimental setup, and data evaluation are provided in the Methods section.

Figure~\ref{fig:fig_2_experiment}b displays the temporal evolution of the normalised diffracted X-ray intensity around the FeRh~(002) Bragg peak along the out-of-plane reciprocal coordinate $q_z$, i.e., the sonogram of FeRh, for an incident optical laser fluence of \qty{7.7}{\mJ\per\cm\squared}. 
Upon optical excitation, we observe the rise of a second Bragg peak at smaller $q_z$ values, corresponding to larger lattice constants, that originates from the emerging FM phase. 
This is a clear signature of the laser-induced magneto-structural phase transition.
At the same time, the optical excitation launches propagating strain pulses that are employed to mark the phase heterogeneity.
The position of the structural Bragg peaks $q_z^{\mathrm{002}}=4\pi/c$ determines the average out-of-plane lattice constant $c$ of both phases.
The strain responses of the AFM and FM phases (see Fig.~\ref{fig:fig_2_experiment}c) are in good approximation given by the relative change of their Bragg peak positions with respect to the values before excitation, $\eta(t) \simeq \frac{q_{z,0}^\mathrm{002}-q_z^\mathrm{002}(t)}{q_{z, 0}^\mathrm{002}}$.
The strain of the FM phase is calculated with respect to the expected equilibrium value and is only shown when the Bragg peak's integral exceeds a certain threshold.
Comparing these strain responses with the simplified scenarios from Fig.~\ref{fig:fig_1_idea} directly highlights two major differences: 
(i) The signatures of the strain pulses do not occur symmetrically around zero strain, since both phases mostly expand due to the heating of FeRh via direct photo-excitation~\cite{mattern2023a,mattern2024a}. 
(ii) The strain responses show more than one minimum and maximum due to the acoustic mismatch between the layers of the heterostructure, which results in a (partial) reflection of the strain pulse at the interfaces.

Although the details of the experimental FeRh sonogram are more complex than sketched in Fig.~\ref{fig:fig_1_idea}, we can still directly exclude the scenario of pure in-plane heterogeneity (Fig.~\ref{fig:fig_1_idea}b) due to the major differences in the strain response of the AFM and FM phase.
Distinguishing between the other two key scenarios of pure out-of-plane and mixed heterogeneities is more subtle and requires a comparison of the experimental X-ray sonogram with simulated data, which can even quantify the transient relative thickness $\drel_\text{FM}(t)$ and in-plane coverage $\Arel_\text{FM}(t)$ of the FM phase via Eq.~\eqref{eq:eq_1_separation}.

\section{Ultrafast domain nucleation in FeRh}

\begin{figure*}[!t]
\centering
\begin{minipage}[t]{0.488\textwidth}\vspace{0pt}
\includegraphics[width=1\textwidth]{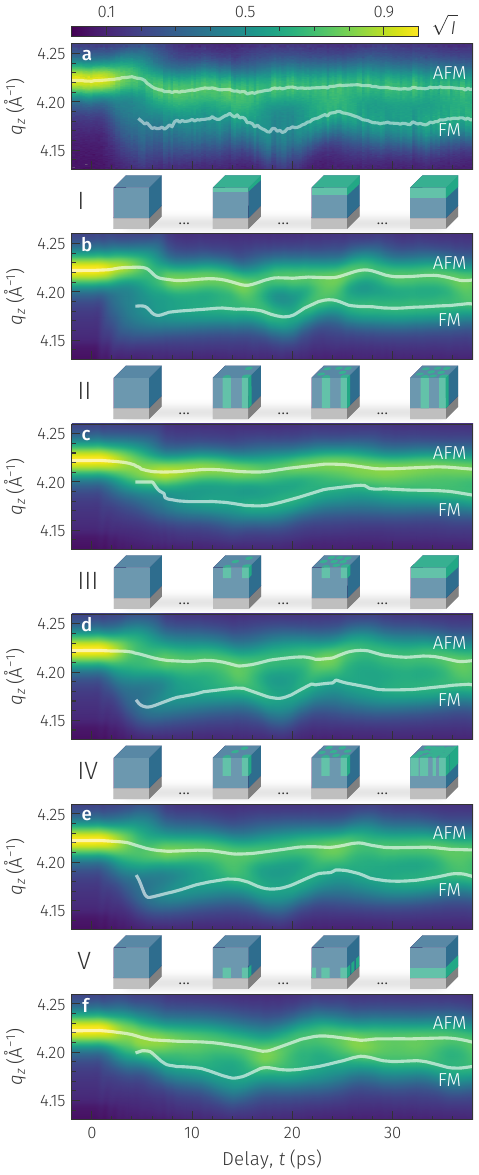}
\end{minipage}%
\hfill
\begin{minipage}[t]{0.488\textwidth}\vspace{0pt}
\captionof{figure}{\label{fig:fig_3_model} \textbf{Comparison of experimental and modelled X-ray sonograms revealing phase heterogeneity:} 
\textbf{a}, Same experimental data as shown in Fig.~\ref{fig:fig_2_experiment}b for comparison.
\textbf{b--f}, Calculated sonograms for the modelled nucleation scenarios, as sketched above each panel:
\textbf{I}, The ferromagnetic (FM) phase nucleates as a continuous layer at the top of the FeRh layer that grows into the depth.
\textbf{II}, The FM phase nucleates as columns through the entire FeRh film that become more dense with progressing nucleation.
\textbf{III}, The FM phase nucleates as columns through the near-surface region of the FeRh layer and finally forms a continuous layer.
\textbf{IV}, The FM phase nucleates as columns in the near-surface region, as in scenario III, but does not form a continuous layer, $\Arel_\text{FM}(t) < 1$.
As an example, we chose a relative thickness $\drel_\text{FM}(t)=1.3 \, \Vrel_\text{FM}^*$ to maintain a fraction at the surface antiferromagnetic (AFM).
\textbf{V}, similar to scenario III, but the FM columns nucleate in the near-substrate region.
In all panels, the solid lines mark the fitted AFM and FM Bragg peak positions.}\noindent
\end{minipage}
\end{figure*}

In the following, we discuss five realistic nucleation scenarios, which are sketched above each panel b--f in Fig.~\ref{fig:fig_3_model}.
The scenarios cover pure out-of-plane~(I), pure in-plane~(II), and mixed-phase heterogeneities~(III-V).
Furthermore, they include different locations of the FM phase within the FeRh layer~(III, V) and a different evolution of the nature of the heterogeneity~(III, IV) associated with a different transient relative thickness $\drel_\text{FM}(t)$ of the nucleated FM phase.
With the help of Eq.~\eqref{eq:eq_1_separation}, we can find simple analytical expressions to specify the individual scenarios, which are described in more detail in the Methods section.
Irrespective of the nucleation scenario, the rise of the FM volume fraction is constrained via Eq.~\eqref{eq:eq_0_model_old} with the fixed nucleation timescale $\tau = \qty{8}{ps}$ \cite{mattern2024a, mattern2024b, mattern2025} and the experimentally determined, final FM volume fraction $\Vrel_\text{FM}^*=0.46$ for an incident optical laser fluence of $\qty{7.7}{\mJ\per\cm\squared}$.

Our model of the sonogram is based on calculating laser-induced stresses and the resulting spatio-temporal strain by solving the linear one-dimensional elastic wave equation~\cite{mattern2023a}.
The essential thermophysical parameters (see Methods section) can be calibrated most easily for excitation scenarios that do not drive the phase transition.
Here, we use essentially the parameters already calibrated in our previous experiment on the very same sample~\cite{mattern2024b}.
In addition, we explicitly consider the unit cell's volumetric changes through the phase transition by introducing additional stresses that correspond to an out-of-plane expansion of \qty{0.6}{\percent} with the appearance of the FM phase~\cite{arregi2020, mattern2024a}.
Finally, we obtain the modelled sonograms in Fig.~\ref{fig:fig_3_model}b--f by feeding the transient strains into a dynamical X-ray scattering formalism~\cite{schick2021}.
For the case of lateral phase coexistence, we account for the stochastic nature of the domain nucleation by calculating incoherent averages of the X-ray intensities for different nucleation delays weighted by the corresponding probabilities.
More details on the modelling are provided in the Methods section.

Qualitatively, we find that the experimental sonogram in Fig.~\ref{fig:fig_3_model}a displays much stronger transient variations of the FM Bragg peak position compared to the AFM counterpart (solid white lines).
Comparing these signatures to the modelled sonograms again excludes the nucleation scenario of pure in-plane heterogeneity (II), exhibiting similar shifts of both peaks (c), and at the same time, the situation of a backside nucleation (V), showing a stronger modulation of the AFM Bragg peak instead (f).

\begin{figure}[t]
\centering
\includegraphics[width = 1\columnwidth]{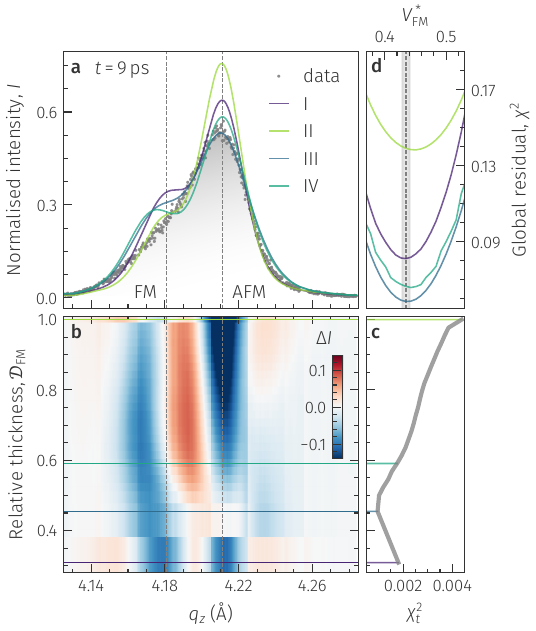}
\caption{\label{fig:fig_4_quantitative} \textbf{Quantitative comparison of experimental and modelled sonograms:}
\textbf{a}, Comparison of the normalised X-ray intensity $I(t=\qty{9}{ps}, q_z)$ for scenarios I--IV (solid lines) with the experiment (symbols).
\textbf{b}, Difference intensities, $\Delta I$, between experiment and model for varying relative thickness of the ferromagnetic (FM) phase $\drel_\text{FM}(t=\qty{9}{ps})$.
The solid horizontal lines denote the scenarios I--IV from Fig.~\ref{fig:fig_3_model}.
The dashed vertical lines mark the experimentally determined positions of the FM and antiferromagnetic (AFM) Bragg peaks.
\textbf{c}, The quadratic deviation $\chi_t^2$ between experiment and model shows a minimum for scenario III at $t=\qty{9}{ps}$.
\textbf{d}, The global quadratic deviation between experiment and model summed over all delays $\chi^2=\sum_t \chi^2_t$.
The vertical dashed line and the grey shaded area denote the experimentally determined $\Vrel_\text{FM}^*=0.46$ and the corresponding standard error, respectively.
}
\end{figure}

To distinguish between the three remaining scenarios (I, III, IV), we quantitatively compare the measured and modelled sonograms.
While the extraction of the relevant Bragg peak parameters (intensity, position, and width) via established peak fitting routines can provide a more detailed view on the nucleation dynamics in FeRh, see Extended Data Figure~\ref{fig:fig_z1_fit}, especially for early delays ($< \qty{10}{ps}$), the complex strain dynamics and the initial domain nucleation result in diffraction patterns that can hardly be described analytically.
To that end, we take advantage of the full information content of the experiment and the simulations by comparing directly their normalised X-ray intensities $I(t, q_z)$, as exemplified for a selected delay of $t=\qty{9}{ps}$ in Fig.~\ref{fig:fig_4_quantitative}a.
At this delay, the expansive part of the bipolar strain pulse starts to exit the FM phase as indicated by the decreasing FM strain in the experiment (see Fig.~\ref{fig:fig_2_experiment}c).
Hence, the average strain, i.e. the FM Bragg peak position, is a sensitive measure of the relative thickness of the FM phase as illustrated by the shift of the simulated FM Bragg peak towards smaller (larger) $q_z$ when increasing (decreasing) $\drel_\text{FM}(t=\qty{9}{ps})$ from scenario III to IV (III to I).
In addition to the four distinct scenarios (solid lines), we plot the difference between the experimental and modelled intensities, $\Delta I$, in Fig.~\ref{fig:fig_4_quantitative}b by systematically varying $\drel_\text{FM}$ while adjusting $\Arel_\text{FM}$ according to Eq.~(\ref{eq:eq_1_separation}).
The corresponding quadratic deviation $\chi_t^2 = \sum_{q_z}{\left|\Delta I(t, q_z)\right|^2}$ averaged over $q_z$ is given in panel c and displays a clear minimum for scenario III at this specific delay $t = \qty{9}{ps}$.
The maximum of $\chi_t^2$ occurs for the pure in-plane heterogeneity (II), which is in line with the exclusion of this scenario based on the qualitative comparison of the phase-specific strain response in Fig.~\ref{fig:fig_3_model}.
At $t=\qty{9}{ps}$ the strong deviation mostly originates from the narrowing of the AFM Bragg peak with increasing $\drel_\text{FM}$, i.e. decreasing in-plane coverage $\Arel_\text{FM}$.
Since the width of the Bragg peak is a measure of the homogeneity of the out-of-plane lattice constant, it decreases when the AFM volume is less inhomogeneously strained, e.g., by distributing the nucleating FM domains strongly straining the adjacent AFM regions over the full FeRh film thickness ($\drel_\text{FM} = 1)$.
This illustrates the importance of a complete peak shape and position analysis to capture the full information on the heterogeneity across volume-changing phase transitions.

In the sonography analysis, we further account for all pump-probe delays, $t$, by minimising the global residual $\chi^2 = \sum_t{\chi_t^2}$.
We display $\chi^2$ as a function of the final FM volume fraction $\Vrel_\text{FM}^*$ to exclude any local minima in Fig.~\ref{fig:fig_4_quantitative}d.
Again, this global analysis identifies scenario III to optimally describe the nucleation of the FM phase during the laser-induced magneto-structural phase transition for the investigated FeRh layer, excited by an optical laser fluence of \qty{7.7}{\mJ\per\cm\squared}.
Importantly, our result is in line with a much slower rise of the macroscopic magnetisation, as shown in previous experiments~\cite{li2022,mariager2012}, governed by aligning the magnetisation of the nucleated FM domains that initially points equally along the different magnetic easy axes via domain wall motion.

To test the generality of our findings for FeRh, we apply the ultrafast X-ray sonography to three additional optical laser fluences covering a broad range of $\Vrel_\text{FM}^*$ (Extended data Figure~\ref{fig:fig_z2_data}) and identify scenario III to describe the nucleation kinetics for fluences $\geq \qty{5.2}{\mJ\per\cm\squared}$ (Extended Data Figure~\ref{fig:fig_z3_deviation}).
Here, the fluence-dependent relative thickness $\drel_\text{FM}$ can be approximated by the depth until which the optical excitation overcomes the equilibrium transition threshold.
This reveals the thermal character of the phase transition even during its non-equilibrium pathway.
For the lowest fluence of \qty{3.9}{\mJ\per\cm\squared}, however, a final in-plane coverage of $\Arel_\text{FM}^*=0.66$ (scenario IV) matches the experiment best.
Considering the lateral heterogeneous transition temperature~\cite{baldasseroni2015,ahn2022}, this is reasonable, as the weak excitation overcomes the transition threshold only for regions exhibiting the lowest transition temperature.
In other words, for low fluences the phase transition remains in-plane heterogeneous, while higher fluences that overcome the transition threshold in the entire X-ray probe footprint transform the entire near-surface region into the FM phase.

In the last step, we estimate the lateral size of the nucleating domains.
In the simplest case of a large intrinsic structural in-plane coherence length of the film, the width of the FM Bragg peak along the in-plane reciprocal coordinates $q_x$ and $q_y$ directly determines the in-plane size of the FM domains~\cite{zeuschner2024}.
However, the investigated FeRh film is composed of small mosaic crystallites with an average diameter of approximately \qty{25}{nm}~\cite{arregi2020} significantly broadening the Bragg peaks along $q_x$ and $q_y$.
Instead, we employ the broadening of the FM Bragg peak along the out-of-plane reciprocal coordinate $q_z$ to estimate the lateral diameter of the nucleated FM domains to approximately \qty{30}{nm} (further details are given in the Methods section and in Extended Data Figure~\ref{fig:fig_z4_in_plane}).
For FeRh, we can exploit the fact that the laterally heterogeneous nucleation of FM domains breaks the symmetry and unlocks an in-plane lattice expansion that is otherwise forbidden~\cite{mattern2023a}.
The in-plane expansion reduces the out-of-plane lattice constant via the Poisson effect.
Thus, the total FM Bragg peaks becomes the superposition of an ensemble of Bragg peaks at different positions representing the individual FM domains nucleated at different delays and sites.
Indeed, the experiment (see Fig.~\ref{fig:fig_4_quantitative}a and Extended Data Figure~\ref{fig:fig_z1_fit}c) shows an additional broadening of the FM Bragg peak that is not captured by our one-dimensional model, disregarding the in-plane expansion.

\section{Discussion \& Outlook}
Our results highlight the power of ultrafast X-ray sonography in identifying the nature of transient directional phase heterogeneity, as well as quantifying the dimensions of coherently or stochastically nucleating domains, which is essential for the functionalisation of materials in technological applications.
Strain pulses, which are typically triggered by the ultrafast driver of the phase transition, do not hinder our method but instead can be readily exploited as structural markers, rendering ultrafast X-ray sonography extremely sample-agnostic.
The matching size of the nucleating FM domains and the mosaic crystal blocks of the FeRh layer points to the crucial role of structural heterogeneity and granularity for the ultrafast emergence of long-range order in this prototypical material.

While real-space techniques, such as time-resolved X-ray imaging, can provide a more detailed picture of ultrafast phase transitions, they are typically limited to material systems where the spatio-temporal dynamics are not stochastic, e.g. due to the presence of defect sites with pronounced pinning~\cite{ahn2022,johnson2023,mcclellan2024}.
In contrast, ultrafast X-ray sonography offers a non-invasive probe of both coherent growth and stochastic nucleation as long as the phases exhibit distinguishable diffraction signatures.
We are confident that even situations in which a phase transition is driven by a strain pulse~\cite{amano2024} can be studied, as long as the coupling is not highly nonlinear.
The effective temporal resolution in ultrafast X-ray sonography is determined by the propagation time of the strain pulses, which is given by the speed of sound and the phase thickness.
This limitation, however, affects only pure symmetry changing transitions as the appearance of a structural phase through volume changes is anyway limited by the speed of sound~\cite{mariette2021}.

To further improve the sensitivity of ultrafast X-ray sonography, we propose double-pulse excitation schemes, in which a first excitation drives the phase transition and a delayed second excitation launches a dedicated strain pulse.
The strain pulse then marks the phase heterogeneity at the selected \emph{pump-pump} delay. 
Utilising a photoacoustic transducer with a band gap enables two-colour double-pulse excitation schemes that selectively excite the transducer without modifying the transition kinetics of the actual sample, and vice versa.
Although the choice of hard X-rays renders sonography very flexible in sample design and allows access also to buried layers~\cite{mattern2023a}, the X-ray photon energy can be tuned to absorption resonances down to the tender X-ray range and potentially even into the soft X-ray range to selectively probe different degrees of freedom and/or elements.
Since the emergence of long-range ordered spins, charges, and orbitals is often coupled to the crystal lattice that parametrises the phase transition, ultrafast X-ray sonography is a universal non-invasive tool providing quantitative access to phase heterogeneity in a broad range of crystalline materials.

\section{Methods}

\subsection{Sample growth and characterisation}
The \qty{44}{nm}-thick FeRh(001) film was grown by magnetron sputtering from an equiatomic FeRh target \cite{arregi2020} on a MgO(001) substrate. It is sandwiched by a $\qty{2}{nm}$-thick Pt capping layer and a $\qty{8}{nm}$-thick W buffer layer. The layer thicknesses were characterised via X-ray reflectometry. It is the very same same sample as in a previous publication \cite{mattern2024b} and it was previously thoroughly statically and structurally characterised \cite{arregi2020}. In thermal equilibrium, the AFM-to-FM phase transition occurs at around $\qty{370}{K}$.

\subsection{UXRD experiment at the MID instrument at EuXFEL}
The experiment was performed at the Materials Imaging and Dynamics (MID) instrument \cite{madsen2021} at the European X-ray Free-Electron Laser Facility (EuXFEL). 

The sample was excited at room temperature (\qty{300}{K}) with ultrashort optical laser pulses with a central wavelength of \qty{800}{nm}.
The laser system provided six \qty{50}{fs} p-polarized laser pulses in a burst with \qty{56.25}{kHz} and a base frequency of \qty{10}{Hz}.
The beam diameters were adjusted to $\qty{136}{\micro m} \times \qty{142}{\micro m}$ (major and minor FWHM diameter).
Under an incident angle of $\approx 25$\textdegree{} to the sample surface, \qty{1}{\micro J} to \qty{3}{\micro J} laser pulses reached the sample resulting in an incident fluence of \qty{3.9}{\mJ\per\cm\squared} to \qty{11.7}{\mJ\per\cm\squared}.

The lattice was probed with \qty{11.2}{keV} X-ray pulses with a bandwidth of $\approx \qty{1}{eV}$ achieved by a hard X-ray self-seeded (HXRSS) operation of the SASE2 undulator of EuXFEL \cite{liu2023,geloni2009}.
11 X-ray pulses with the double burst frequency of the laser system (\qty{112.5}{kHz}) were used to obtain unpumped reference images between the pumped images.
The X-rays were collimated to \qty{50}{\micro m} on the sample using local nanofocusing optics, i.e.\ a stack of 20 Be compound refractive lenses with a focal length of \qty{460}{mm}, positioned \qty{385}{mm} before the sample.
The scattering of a Kapton target behind the lens stack was measured by a Jungfrau detector as a reference signal for the incoming X-ray flux on the sample. 
The FeRh sample was placed at a \qty{20.9}{\degree} incident angle of the X-rays in a horizontal specular geometry in between of the maxima of the AFM Bragg peak at \qty{21.76}{\degree} and the FM Bragg peak at \qty{22.04}{\degree}.
The adaptive gain integrating pixel detector (AGIPD) \cite{sztuk-dambietz_operational_2024} was positioned at \qty{3.8}{m} distance to the sample such that both Bragg peaks can be observed in one quadrant of the detector.
The detected charges from both detectors (AGIPD and Jungfrau) were converted into photon counts during calibration. 
For data evaluation only data were selected if the high resolution hard X-ray single-shot spectrometer (HIREX) detected a decent self-seeding intensity above a threshold ($>60\%$ of the mean self-seeding intensity) and the intensity of the normalisation detector (Jungfrau) is above $2/3$ of its mean value.
The Jungfrau detector data is integrated along the image axis to receive a pulse and train-resolved normalisation intensity $i_0$. The normalised detector images from the AGIPD are integrated along the vertical direction and the with $i_0$ weighted average for one delay is calculated.
Two pseudo-Voigt profiles are fitted to the projected curves to extract the intensity $I$, position $q_z^\text{002}$, and width $w$ of the AFM and the FM Bragg peaks, respectively.
To achieve smaller intensity fluctuations of the presented data the intensity of the pumped data was scaled by the unpumped reference data recorded within the same pulse train.

\subsection{Modelling of sonograms}
We model the time-dependent X-ray diffraction, i.e. the sonogram, using the modular \textsc{Python} library \textsc{udkm1Dsim} \cite{schick2021}.
In the first step, we calculated the spatio-temporal lattice strain within the FeRh layer.
This procedure is described in detail in a recent review~\cite{mattern2023a}.
In a previous study \cite{mattern2024a}, we measured and modelled the laser-induced strain response of both the FeRh and the W buffer layer for a weak excitation that is not able to drive the phase transition.
This yielded a reliable set of thermophysical parameters to describe the spatio-temporal strain within FeRh upon laser excitation. In this work, we use essentially these already calibrated parameters \cite{mattern2024a,mattern2025}.
Only the optical penetration depth is optimised to account for the different incident angle of the optical laser pulses.
In the framework of our model, we calculated the spatio-temporal distribution of optically deposited energy within the metallic heterostructure in the framework of a diffusive two-temperature model \cite{mattern2023a}. 
The spatio-temporal energy density stored in each subsystem then linearly translates to spatio-temporal stress contributions via subsystem- and material-specific Gr\"uneisen parameters.
Finally, their superposition drives the strain response of the heterostructure according to the linear one-dimensional elastic wave equation that is represented by a linear chain of masses and springs in the framework of the \textsc{udkm1Dsim} library.
Note, for an overall better agreement we assumed that the bottommost \qty{5.5}{nm} of the nominal FeRh layer does not contribute to its Bragg peak, which is probably the result of an interface diffusion of FeRh and W~\cite{arregi2020} due to too strong laser excitations at a high-repetition rate.

In the second step, we use the determined spatio-temporal strain as input for dynamical X-ray diffraction simulations that calculate a layer-specific Bragg peak in reciprocal space for each pump-probe delay.
Additionally, we have to explicitly consider the volume changing character of the phase transition in FeRh that affects the spatio-temporal strain.
Therefore, we describe the nucleation of a FM domain by the appearance of an expansive stress that corresponds to an expansion of \qty{0.6}{\percent}.
Considering the parametrizing of the phase heterogeneity by the in-plane coverage $\Arel_\text{FM}(t)$ and the relative thickness $\drel_\text{FM}(t)$ we can formulate simple analytic constrains for the simulation of the scenarios in Fig.~\ref{fig:fig_3_model}:
In case of a pure out-of-plane heterogeneity (I) the relative thickness is directly given by the transient FM volume fraction $\Vrel_\text{FM}(t)$:  $\Arel_\text{FM}(t)\overset{!}{=}1 \rightarrow \drel_\text{FM}(t)=\Vrel_\text{FM}(t)$.
In case of a pure in-plane heterogeneity (II) the in-plane coverage is directly given by $\Vrel_\text{FM}(t)$: $\drel_\text{FM}(t)\overset{!}{=}1 \rightarrow \Arel_\text{FM}(t)=\Vrel_\text{FM}(t)$.
In case of a mixed heterogeneity finally resulting in the formation of continuos FM layer (III and V) $\drel_\text{FM}(t)$ is directly given by the final FM volume fraction $\Vrel_\text{FM}^*$: $\drel_\text{FM}(t)\overset{!}{=}\Vrel_\text{FM}^* \rightarrow \Arel_\text{FM}(t)=\Vrel_\text{FM}(t)/\Vrel_\text{FM}^*$.
For scenario IV, where no laterally continuous layer is formed, we choose: $\drel_\text{FM}(t)=1.3\Vrel_\text{FM}^* \rightarrow \Arel_\text{FM}(t)=\Vrel_\text{FM}(t)/1.3\Vrel_\text{FM}^*$.

In case of a (transient) lateral heterogeneity (II--V), we additionally implement the stochastic nature of the domain nucleation and calculate the incoherent average of the sonograms for different nucleation delays $t_\text{nuc}$ weighted by the probability according to Eq.~\eqref{eq:eq_0_model_old}.
In case of a non-vanishing AFM in-plane coverage ($\Arel_\text{AFM}(t)>0$), we additionally calculate the incoherent average of this final sonogram for sites with FM domain nucleation with a sonogram for the AFM FeRh layer weighted by the final in-plane coverage $\Arel_\text{FM/AFM}^*$.

While we consider the effect of the nucleating FM domains on the spatio-temporal strain, we do not account for potential minor effects like the previously reported rapid modification of the electronic band structure \cite{pressacco2021} and the potentially associated change of the electronic heat capacity and electron-phonon coupling strength or the reduced thermal expansion in the FM phase and the latent heat that is consumed by the phase transition.
Neglecting these minor effects drastically reduces the free parameters of our model and demonstrates that the identification of heterogeneities of laser-induced structural phase transitions via modelling the phase-specific strain does not depend decisively on the details of the shape of the strain pulse.

\subsection{Estimating the in-plane domain dimension}
Extended Data Figure~\ref{fig:fig_z1_fit}c displays an additional broadening of the FM Bragg peak that is not captured by our modelling.
This additional broadening decays approximately on the nucleation timescale of the FM phase indicating an origin from the stochastic lateral domain nucleation.
In typical pump-probe experiments, the large footprint of the pump pulse compared to the footprint of the probe pulse ensures a laterally homogeneous excitation of the probed volume.
This corresponds to a vanishing total in-plane stress.
Thus, the laser-induced expansion for continuous thin films is limited to the out-of-plane direction \cite{mattern2023a}.
However, the lateral heterogeneous nucleation of the FM domains breaks this symmetry and unlocks an additional in-plane expansion due to the absence of a neighbouring volume element with an identical in-plane stress on the lattice.

A single FM domain with a diameter of $l_\text{FM}$ within the AFM matrix exhibits an in-plane lattice expansion of the unit cell that starts at domain boundary and propagates at the speed of sound $v_s$ through the nucleated domain until it is fully established at delays $>l_\text{FM}/v_s$.
When a FM domain nucleates at $t_\text{nuc}$ at one of the four neighbouring sites, it expands by itself.
This means that the in-plane expansion at the initial site along this direction is erased starting at the domain boundary until $1/4$ of the in-plane expansion has vanished at $l_\text{FM}/v_s+t_\text{nuc}$.
At this delay the compression compensating the initial expansion has propagated through the entire domain.
This process is illustrated in Extended Data Figure~\ref{fig:fig_z4_in_plane}a for $l_\text{FM}=\qty{30}{nm}$ and selected nucleation delays $t_\text{nuc}=\qty{0}{ps}$, \qty{2}{ps}, \qty{4}{ps}, \qty{8}{ps} and \qty{20}{ps} for one nucleation site and its four neighbouring nucleation sites.
In the next step, we explicitly include the stochastic nature of the domain nucleation by averaging situations as sketched in Extended Data Figure~\ref{fig:fig_z4_in_plane}a for different $t_\text{nuc}$ weighted by their probability to match the rise of the FM volume fraction according to Eq.~\eqref{eq:eq_0_model_old}.

Extended Data Figure~\ref{fig:fig_z4_in_plane}b displays the resulting average transient in-plane strain of the five nucleation sites.
The different amplitude and sign of the in-plane strain among the nucleation sites correspond to different out-of-plane lattice constants via the Poisson effect.
Thus, each site would exhibit its own Bragg peak at different positions along the out-of-plane reciprocal coordinate $q_z$.
This translates to a broadening of the total Bragg peak given by their superposition.
We match the additional broadening of the FM Bragg peak by \qty{0.06}{\per\angstrom} in Extended Data Figure~\ref{fig:fig_z1_fit}c compared to our model employing the \textsc{udkm1Dsim} toolbox~\cite{schick2021} if we assume the sites to have a size of \qty{30}{nm}.
If the domains would be much smaller, the amplitude of the average in-plane strain would be much larger since the in-plane expansion could be established in a larger fraction of the volume before it is reversed by the nucleation at the neighbouring sites.
If the domains would be much larger, the amplitude of the average in-plane strain would be much smaller since the in-plane expansion could only be established in a small fraction of the volume before it is reversed by the nucleation at the neighbouring sites.
Considering the uncertainty whether the additional broadening in the experiment exclusively originates from the broken in-plane symmetry of the stress and the simplification of not explicitly modelling a large network of nucleation sites but only a fundamental building block of $5$ nucleation sites, we estimate the uncertainty of our in-plane determination to \qty{10}{nm}.

\bibliography{references.bib}

\section*{Acknowledgements}
We acknowledge European XFEL in Schenefeld, Germany, for the provision of XFEL laser beamtime at the Materials Imaging and Dynamics instrument located at the SASE2 beamline under proposal number p900401 and would like to thank the staff for their assistance.
This work was supported by the computational resources of the Maxwell at Deutsches Elektronen-Synchrotron (DESY), Hamburg, Germany.

\section*{Funding}
MM and DS thank the Leibniz Association for funding through the Leibniz Junior Research Group J134/2022.
JAA and VU acknowledge Project no. CZ.02.01.01/00/22\_008/0004594 (TERAFIT). Access to the CEITEC Nano Research Infrastructure was supported by the Ministry of Education, Youth and Sports (MEYS) of the Czech Republic under the project CzechNanoLab (LM2023051).

\section*{Data availability}
The data is available from the authors upon request.

\section*{Code availability}
The codes used to evaluate the data and simulate the dynamics are available from the authors upon request.

\section*{Author contributions}
JEP, MM, DS supervised and coordinated the project. JEP conceived and conducted the experiment at MID. The setup was prepared by ARF, RS, JEP, AM. The MID instrument was operated by JEP, ARF, RR, WJ, UB, JH, AL, RR, AZ. JAA, VU produced and characterised the sample. JEP and MM performed the data evaluation with help of JW. MM conducted the simulations and modelling for which JEP and DS added theoretical background. MM and DS wrote the manuscript with the help of JEP. All authors discussed the results and contributed to the manuscript.

\clearpage
\onecolumn
\begin{appendices}

\newcounter{efigure}
\setcounter{efigure}{5}
\renewcommand{\theefigure}{\Alph{efigure}}
\counterwithin{figure}{efigure}
\renewcommand{\thefigure}{\theefigure\arabic{figure}}

\section*{Extended Data}
\renewcommand{\figurename}{Extended Data Fig.}

\begin{figure}[htb]
\centering
\includegraphics[width = 0.5\textwidth]{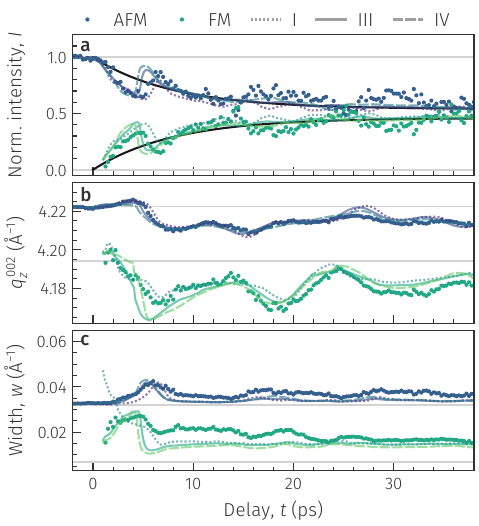}
\caption{\textbf{Comparison of Bragg peak properties in experiment and model:} 
\textbf{a}, The normalised intensity $I$ of the fitted antiferromagnetic (AFM) and ferromagnetic (FM) Bragg peaks (symbols) as a function of delay $t$. 
The black solid line denotes the expected phase transformation on an \qty{8}{ps} timescale (cf. Eq.~\eqref{eq:eq_0_model_old}). 
The green and blue lines denote the evolution of both Bragg peaks for the nucleation scenarios I, III, and IV.
\textbf{b}, \textbf{c}, Peak position $q_z^\text{002}$ and peak width $w$ as a function of delay $t$. 
The peak widths are offset for clarity, and the light-grey solid lines denote the values before excitation.}
\label{fig:fig_z1_fit}
\end{figure}

\begin{figure*}[htb]
\centering
\includegraphics[width = 0.5\textwidth]{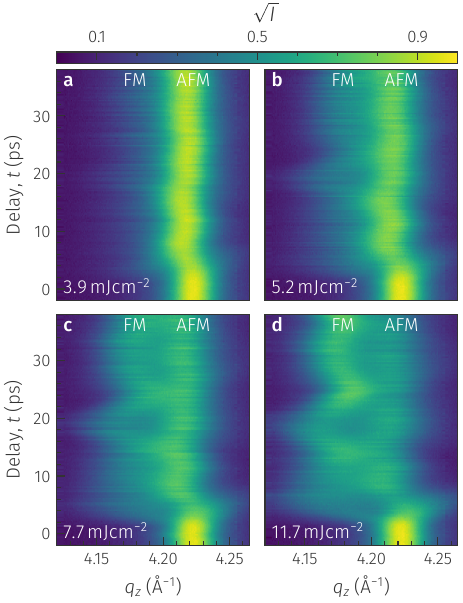}
\caption{\textbf{Fluence-dependent sonograms:} The normalised diffracted X-ray intensity $I$ as function of the out-of-plane reciprocal coordinate $q_z$.
The position of the antiferromagnetic (AFM) and ferromagnetic (FM) structural Bragg peak $q_{z,\text{AFM,FM}}$ quantifies the average out-of-plane lattice constant $c_\text{AFM/FM}=2\pi/q_{z,\text{AFM/FM}}^\text{002}$ of both phases.
Panels \textbf{a}--\textbf{d} display the phase-specific lattice response for increasing optical laser fluences that are above the threshold of the phase transition in FeRh.
Within increasing fluence, the final integral of the emerging FM Bragg peak increases as well as the amplitude of the transient shift of the Bragg peaks, i.e., the thermo-elastic lattice response.
}
\label{fig:fig_z2_data}
\end{figure*}

\begin{figure*}[htb]
\centering
\includegraphics[width = \textwidth]{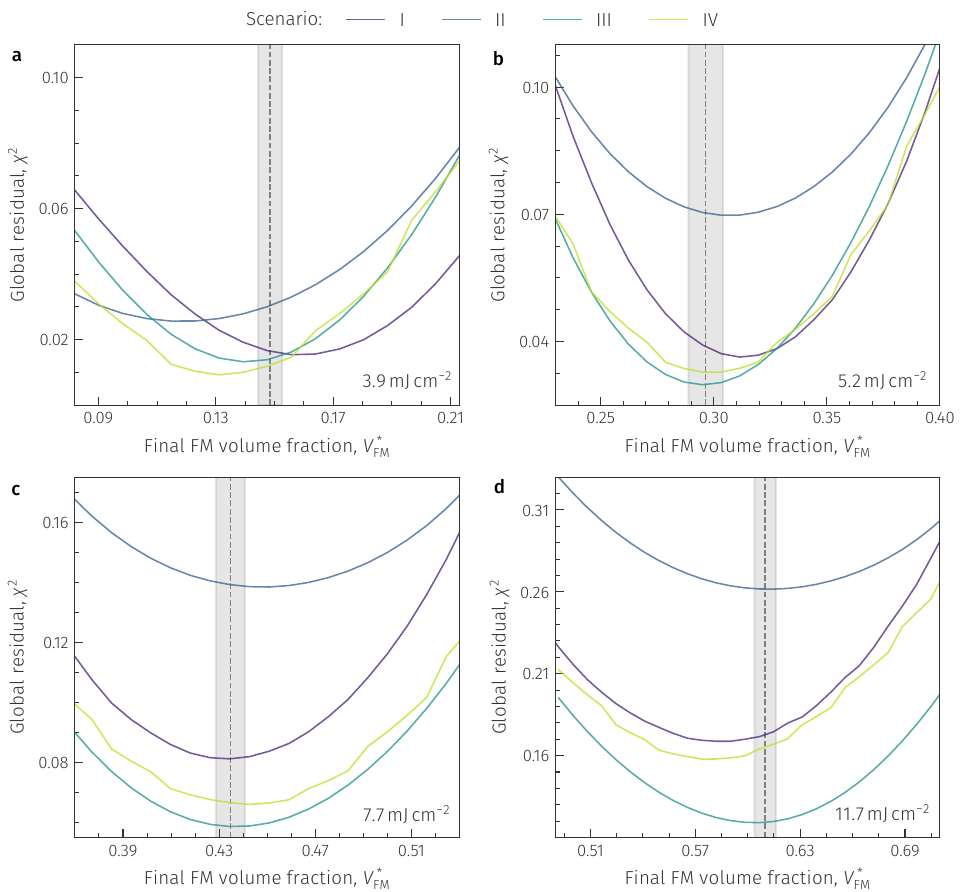}
\caption{\textbf{Quantitative analysis deviation model and experiment:} 
The global residual $\chi^2$ as function of the final ferromagentic (FM) volume fraction $\Vrel_\text{FM}^*$ for the nucleation scenarios I, II, III and IV at 4 different laser fluences \textbf{a} 3.9, \textbf{b} 5.2, \textbf{c} 7.7 and \textbf{d} \qty{11.7}{\mJ\per\cm\squared}.
The deviation of scenario V is even larger as scenario II and is therefore omitted.
The vertical dashed line and the gray shaded area denote the experimentally determined $\Vrel_\text{FM}^*$ and the corresponding standard error, respectively.}
\label{fig:fig_z3_deviation} 
\end{figure*}

\begin{figure*}[htb]
\centering
\includegraphics[width = \textwidth]{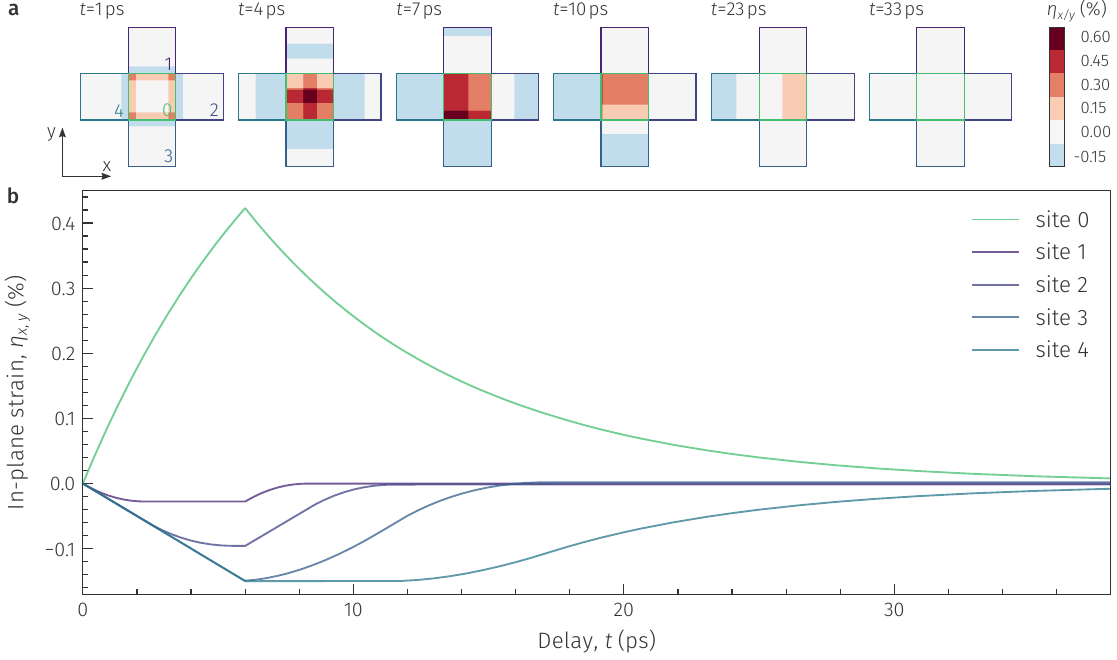}
\caption{\textbf{Model of the transient in-plane expansion:} 
\textbf{a} Sketch of the spatio-temporal in-plane expansion $\eta_{x/y}$ that evolves with the speed of sound through the five neighbouring nucleation sites of \qty{30}{nm} size in case of the nucleation at $t_\text{nuc}= \qty{0}{ps}$, \qty{2}{ps}, \qty{4}{ps}, \qty{8}{ps} and \qty{20}{ps} at site $0$, $1$, $2$, $3$ and $4$, respectively. 
\textbf{b} The corresponding average in-plane strain of the five nucleation sites including the stochastic nature of the nucleation process by averaging the in \textbf{a} sketched dynamics for different $t_\text{nuc}$ weighted by the probability given by Eq.~\eqref{eq:eq_0_model_old}.
This spread in the in-plane expansion translates to a transient spread of the out-of-plane lattice constant via the Poisson effect.
This transient spread in the out-of-plane lattice constant explains the additional broadening of the FM Bragg peak in the experiment, cf. Fig.~\ref{fig:fig_4_quantitative}a and Extended Data Figure~\ref{fig:fig_z1_fit}.}
\label{fig:fig_z4_in_plane}
\end{figure*}

\end{appendices}
\end{document}